\documentclass[12pt,preprint]{aastex}

\shorttitle{Dutra Stellar Clusters} \shortauthors{Zhu et al.}

\begin{document}

\title{A Near-Infrared Study of the Stellar Cluster: [DBS2003]~45}
\author{QINGFENG ZHU\altaffilmark{1}, BEN DAVIES\altaffilmark{2,4}, DONALD F. FIGER\altaffilmark{3}, CHRISTINE TROMBLEY \altaffilmark{3}}

\altaffiltext{1}{Key Laboratory for Researches in Galaxies and
Cosmology, the Chinese Academy of Sciences, Center for Astrophysics,
University of Science and Technology of China, 96 Jinzhai Rd.,
Hefei, Anhui, 230026, China; zhuqf@ustc.edu.cn}

\altaffiltext{2}{School of Physics and Astronomy, The University of
Leeds, Woodhouse Lane, Leeds LS2 9JT, United Kingdom;
b.davies@leeds.ac.uk}

\altaffiltext{3}{Chester F. Carlson Center for Imaging Science,
Rochester Institute of Technology, 54 Lomb Memorial Drive,
Rochester, NY 14623-5604; figer@cis.rit.edu; cmtpci@cis.rit.edu}

\altaffiltext{4}{Visiting Astronomer of the NTT}

\begin{abstract}
We present a multi-wavelength photometric and spectroscopic study of
a newly discovered candidate cluster [DBS2003]~45. Our H, Ks
photometry confirms that [DBS2003]~45 is a cluster. An average
visual extinction A$_V$$\sim$7.1$\pm$0.5 is needed to fit the
cluster sequence with a model isochrone. Low resolution spectroscopy
indicates that half a dozen early B and at least one late O type
giant stars are present in the cluster. We estimate the age of the
cluster to be between 5 and 8~Myr based on spectroscopic analysis.
Assuming an age of 6~Myr, we fit the observed mass function with a
power law, N(M)~$\varpropto$~M$^{-\Gamma}$, and find an index
$\Gamma\sim$~1.27$\pm$0.15, which is consistent with the Salpeter
value. We estimate the total cluster mass is around 10$^3$M$_\odot$
by integrating the derived mass function between 0.5 and
45~M$_\odot$. Both mid-infrared and radio wavelength observations
show that a bubble filled with ionized gas is associated with the
cluster. The total ionizing photon flux estimated from radio
continuum measurements is consistent with the number of hot stars we
detected. Infrared bright point sources along the rim of the bubble
suggest that there is triggered star formation at the periphery of
the HII region.
\end{abstract}


\section{Introduction}
Stellar clusters serve as ideal laboratories to study the evolution
of stars of different masses, as the stars in a cluster are formed
at the same time from the same molecular cloud. This assumption
predicts that the stars are at the same distance, have practically
the same initial chemical abundances, and evolve in a similar
physical environment. Such a coeval population can be used to make
critical tests of theoretical stellar evolution models. Young
clusters are of particular interest for this purpose because most of
their massive stellar members are still in main sequence (MS) or
post-MS stages before supernova explosions. Studies of such stars
can improve our understanding about the formation and evolution of
massive stars, as well as the evolution of whole clusters in
general.

Stars generally form with a frequency that decreases with increasing
mass for M$>$1 M$_{\odot}$, i.e. $\Gamma$=d(log N)/d(log m) $\sim$
-1.35 \citep{sal55,sca98,kro01,kro02}. This initial mass function
(IMF) has been used extensively. However, some evidence suggests
that the function may not be universal. \cite{figKMS99},
\cite{stoGBF02} and \cite{kimFKN06} claim a significantly shallower
slope for the Arches cluster. This result, for one of the most
massive and the densest cluster in the Galaxy, suggests that more
massive and dense clusters have a shallow IMF. Alternatively,
perhaps the Galactic center favors shallow IMFs \citep{mor93}. Since
the existing IMF slope of -1.35 was first measured by \cite{sal55}
for field stars with masses lower than 10~M$_\odot$ and field stars
have a MF which is different from the IMF of clusters, the claim of
a universal IMF seems questionable. \cite{zinY07} commented that,
although massive stars can form locally in the field, evidence
suggested that massive stars form preferentially in dense clusters
or OB associations. Additionally, mass segregation can cause massive
stars to sink toward the cluster center and the cluster continuously
loses low mass stars to the field. These effects would modify the
cluster MF so that it becomes shallower as the age of the cluster
increases. The ultimate resolution to this question will come from
measurements of the IMF in stellar clusters throughout the Galaxy.
The sample of newly-identified clusters provides a fresh opportunity
for this purpose.

There are less than 1600 open clusters discovered at optical
wavelengths in our galaxy \citep{diaAML02}. It is reasonable to
believe that a large number of such clusters are missed by optical
observations due to substantial line-of-sight interstellar
extinction in the Galactic plane. The advance of infrared technology
allows surveys to be carried out in near infrared (NIR) and mid
infrared (mid-IR) wavelengths to circumvent the problem of high
extinction.
Systematic searches for Galactic open cluster candidates have been
carried out in infrared bands and more than 1000 new candidates were
discovered \citep{bicDSB03,dutBSB03,dutOBBZ03,froSR07}, although
many of them remain unconfirmed and extensive efforts in follow-up
observations are needed to determine their true identities..

Here we present our photometric and spectroscopic analysis of
[DBS2003]~45, a candidate cluster chosen from such an infrared
cluster catalog \citep{dutBSB03}. This target is located at (l, b) =
(283.$^{\circ}$88, -0.$^{\circ}$91) and is about 3.5 degrees on the
sky to the west of the Carina nebula (NGC3372), suggesting that the
candidate cluster is likely located in the Sagittarius-Carina spiral
arm.
In \S~{\bf 2}, we describe the origin of data that are used in this
work. Our data reduction procedure is described in \S~{\bf 3}. In
\S~{\bf 4}, we present the results of our analysis. Some issues
related to our observation are discussed in \S~{\bf 5}. Finally, we
summarize our findings in \S~{\bf 6}.

\section{Observations}

Photometric observations of [DBS2003]~45
were carried out on the ESO 3.8m New Technology Telescope (NTT) with
SofI in 2008 February (ESO proposal ID 080.D-0470 ). SofI is an
infrared spectrograph and imaging camera equipped with a Rockwell
Hg:Cd:Te 1024 $\times$ 1024 HAWAII array of 18.5 micron pixel-size
\citep{mooCL98}. It takes images with a plate scale of
0.288$''$/pixel.
H and Ks band images of the target field were obtained in a dither
pattern in which candidate clusters were shifted to one of four
corners of the detector each time. We used discrete integration
times (DITs) of 2s, coadded 10 times, in 4 dithers. Multiple images
were taken at each dither location, resulting in a total 160s of
integration time for the location of the cluster at each waveband.

We also obtained low-resolution spectroscopy of several stars in the
field of the cluster. We used the `Red' grism in conjunction with
the 0.6\arcsec\ slit, giving a spectral resolution of $R \sim 900$
in the wavelength range 1.9 - 2.4$\mu$m. We employed DITs of 40-60s,
repeated twice at each position. After each integration the target
was nodded 20\arcsec\ along the slit. Each object was observed in 8
nod positions, hence total integration times were 16mins per object.
For fainter objects ($Ks > 10$) we increased the number of nod
positions to 16 in order to improve signal-to-noise. To correct for
atmospheric absorption, the solar-type star Hip059532 was observed
immediately after observing the cluster stars. Dark-frames and
flat-field observations were made during twilight. For wavelength
calibration purposes a neon arc lamp was observed.

Additionally, images of the field were taken in four bands at 3.6,
4.5, 5.8, and 8~$\mu$m on February 2007 using the infrared camera -
IRAC on the Spitzer space telescope (Program ID: 30734). During the
observations, we used the 5 position Gaussian dithering pattern. The
resulting images of each band cover an area of 5$'$ by 5$'$ with a
pixel size of 1.2$''$ in both directions of RA and DEC. The total
integration time on target is 471s.

In order to study the nebulosity surrounding the cluster and its
association with the cluster, we use H$_\alpha$, mid-infrared maps,
and radio survey data. H$_\alpha$ line observations were taken from
superCOSMOS survey archive \citep{parPPH05}. Radio flux is obtained
from Parkes-MIT-NRAO 4850 MHz southern sky survey catalog
\citep{wriGBE94}. We also obtained from the archive an 8~$\mu$m
image of a large area including our field taken by Spitzer/IRAC
(Program ID: 40791).
The mid-IR image and the H$_\alpha$ emission map will be compared in
order to address the association between the cluster and the bubble.

\section{Data Reduction and Analysis}
Our photometric and spectroscopic data sets come from different
sources. Therefore, the reduction procedure is slightly different
for individual data set. The SofI data set includes both photometric
and spectroscopic observations. Here we describe their reduction
procedures separately.

\subsection{SofI Data}

\subsubsection{Photometry}
\begin{figure}
\epsscale{1.0} \plotone{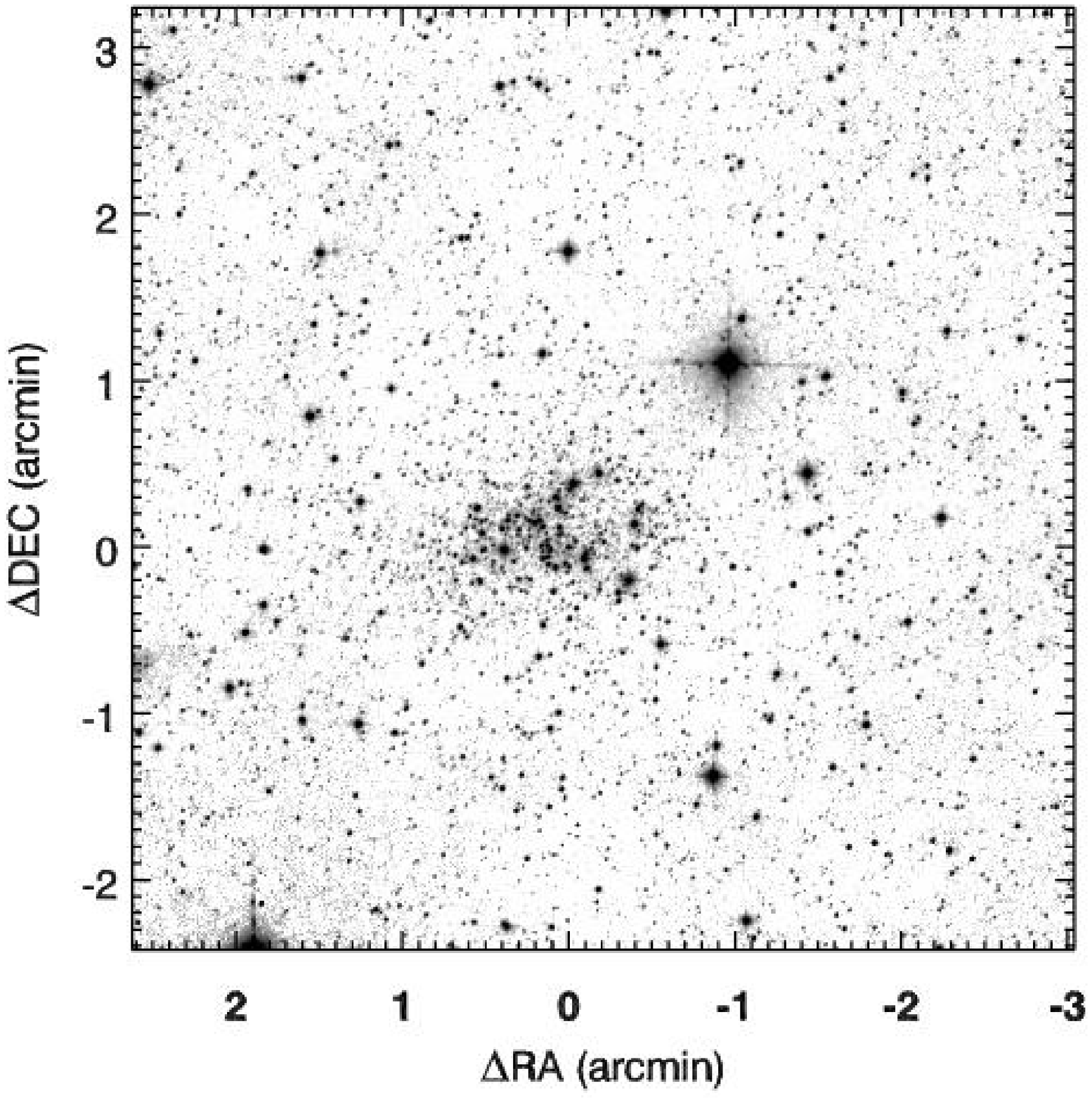} \caption{Ks band mosaic of the
cluster candidate [DBS2003]~45 in a logarithmic scale. The origin of
the axes corresponds to R.A.=10$^h$19$^m$10.5$^s$,
Decl.=-58$^{\circ}$02$'$22.6$''$ (2000). \label{fig_mosaic}}
\end{figure}

Our SofI photometric data consists of two bands, H and Ks, with
eight partially overlapping images in each band. The flat-fielding
is done using facility provided dark and flat frames. A group of 8
images that were obtained closest in time are scaled according to
the integration time and median-combined to create a sky background
image. This background image is scaled and subtracted from each
flat-fielded image of the target. The resulting images are
re-sampled, cross-correlated, shifted, and coadded to create a
mosaic. We register resulting mosaics with downloaded 2MASS mosaics
of the same field in the same photometrical bands to correct for
astrometric errors in our observations.
We manually pick a group of relatively bright and isolated stars
that can be seen in both SofI and 2MASS mosaics. Using the 2MASS
coordinates of these bright stars, we compute the geometric
transformation coefficients between the coordinates of observed
images and the 2MASS images. The observed images are re-mapped onto
the 2MASS grid to produce the final images for photometric analysis.
The final Ks band image is shown in Figure~\ref{fig_mosaic}.

\subsubsection{Calibration}
\begin{figure}  
\epsscale{1.0} \plottwo{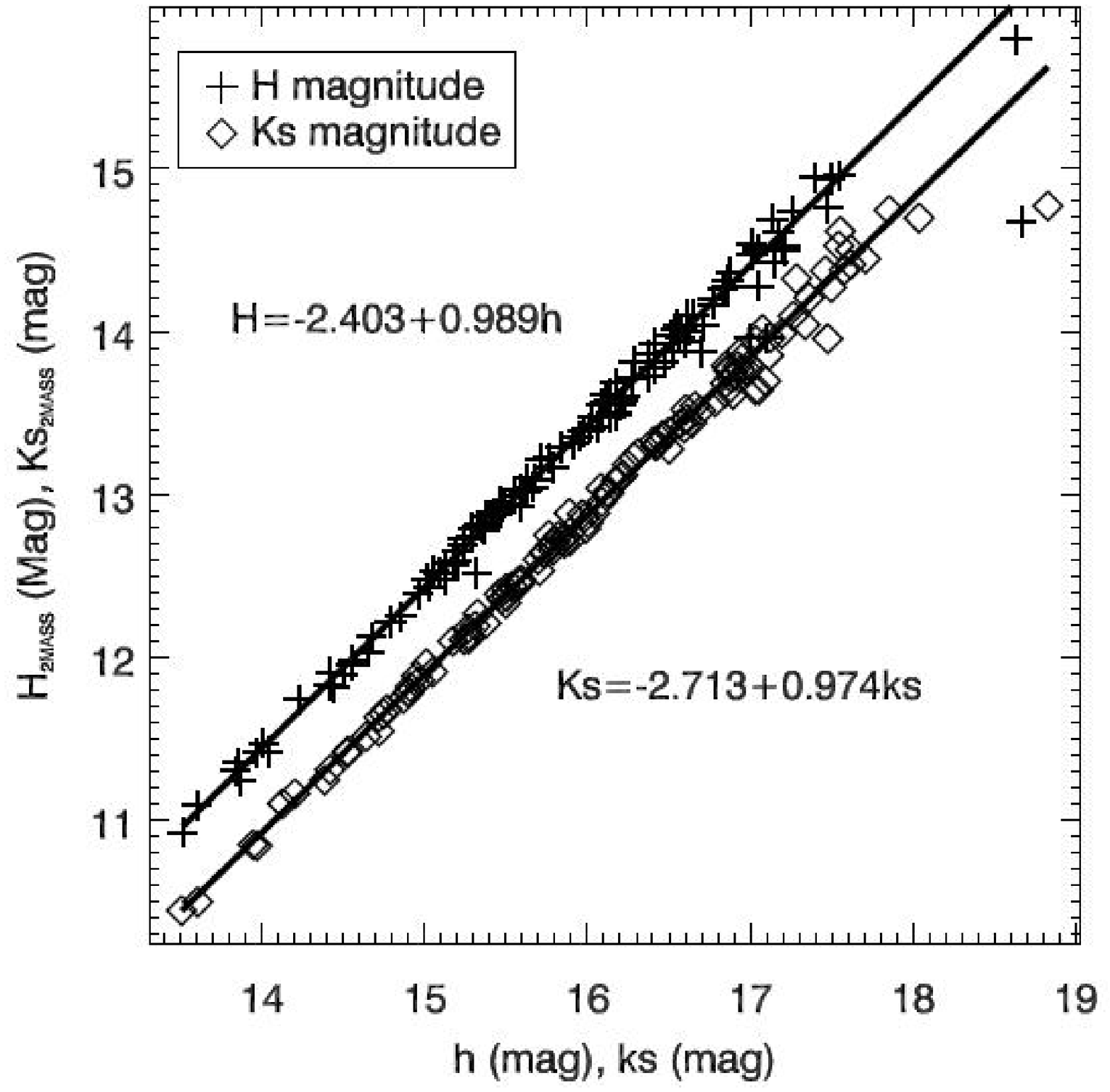}{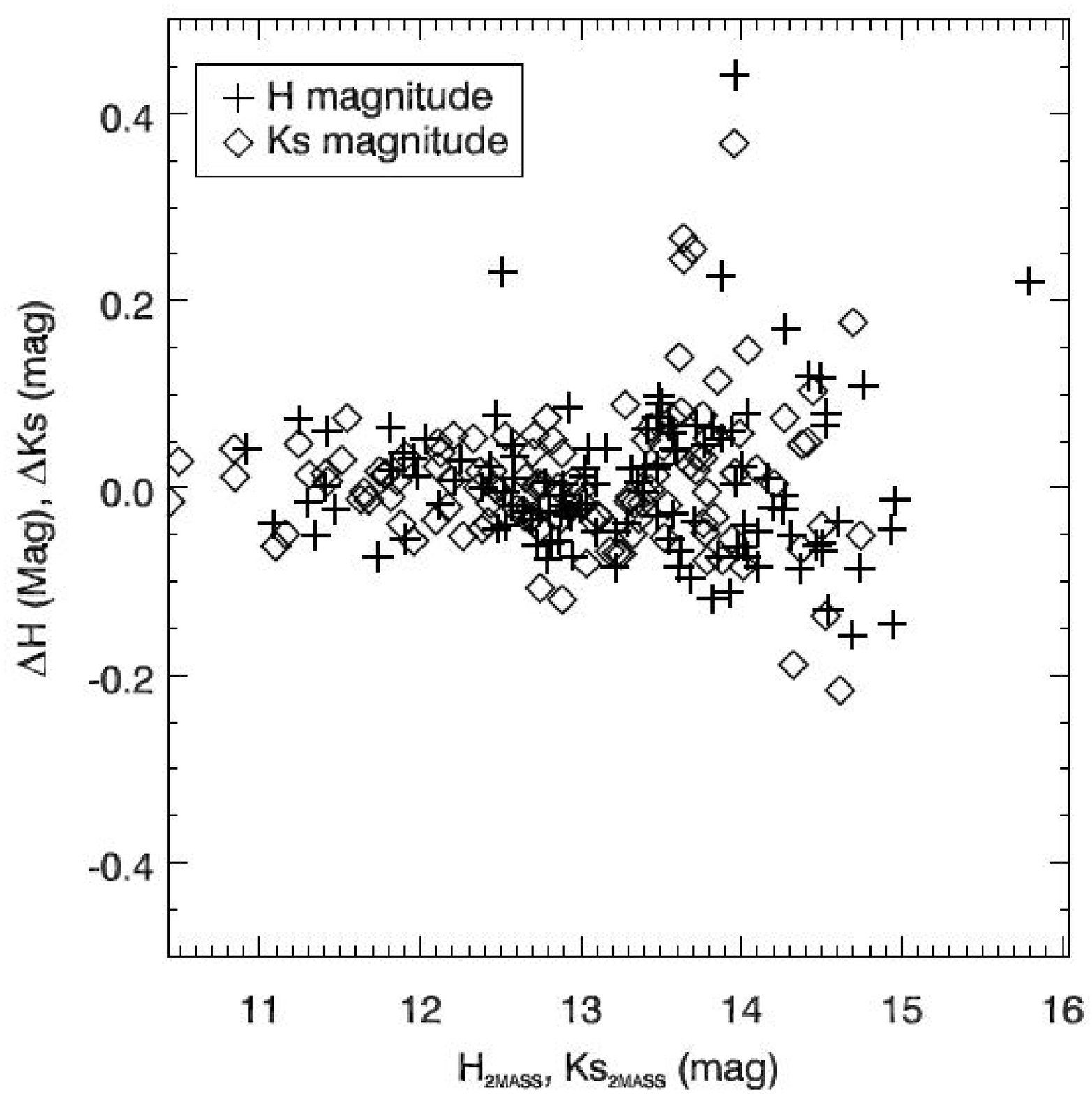} \caption{Using 2MASS
magnitudes to calibrate the observed photometry. In both figures,
plus signs indicate H band magnitudes and diamonds in the figures
indicate Ks band magnitudes. In the left figure, x and y axes are
observed and 2MASS magnitudes, respectively. Solid lines and
equations show the results of a single variable linear model
fitting. In the right figure, x axis is the calibrated magnitude and
y axis is the difference between the calibrated and the instrumental
magnitudes. \label{fig_cali}}
\end{figure}

The aperture photometry of point sources are obtained for the SofI
data using an IDL adapted version of the DAOPHOT. Since all images
are re-mapped onto the 2MASS grid, the magnitudes of different bands
can be associated using stellar locations. The same group of stars
that were used to register the SofI mosaics are used to calibrate
detected H and Ks band magnitudes. Figure~\ref{fig_cali}(left) shows
corresponding instrumental and 2MASS magnitudes of these stars with
plus signs and diamonds representing H and Ks band magnitudes,
respectively. It clearly shows that linear relations (solid lines)
exist between two sets of measurements.

To account for any color term effect present in the data, the
following linear bivariate model between the instrumental magnitudes
and the 2MASS magnitudes is assumed:
\begin{eqnarray}
\label{eq1}
h&=&a_h+b_{h}H+c_{h}Ks\\
ks&=&a_k+b_{k}H+c_{k}Ks
\end{eqnarray}
here, h and ks are instrumental magnitudes of the picked stars. H
and Ks are the corresponding 2MASS magnitudes. The model
coefficients are determined using a regression method. Once all
coefficients have been determined, the following equations are used
to compute calibrated magnitudes of all stars in the field£º
\begin{eqnarray}
\label{eq2}
H&=&\frac{c_{k}h-c_{h}ks-a_{h}c_{k}+a_{k}c_{h}}{b_{h}c_{k}-c_{h}b_{k}}\\
Ks&=&\frac{b_{k}h-b_{h}ks-a_{h}b_{k}+a_{k}b_{h}}{c_{h}b_{k}-b_{h}c_{k}}
\end{eqnarray}
or
\begin{eqnarray}
\label{eq3}
H&=&A_{H}+B_{H}h+C_{H}ks \\
Ks&=&A_{K}+B_{K}h+C_{K}ks
\end{eqnarray}
here, coefficients in the capital form (A, B, C) are the results of
combining the corresponding terms in the equation~\ref{eq2}.
Table~\ref{tab_cali} shows the coefficients found by our regression
procedure and indicates small amount of color effect.
Figure~\ref{fig_cali} (right) shows the corrections between
instrumental and calibrated magnitudes. Again, diamonds and plus
signs in the figure indicate Ks and H band magnitudes, respectively.

\subsubsection{Completeness Test}
The ability to detect stars is impaired when dealing with crowded
fields. Faint stars are likely missed by a star finding procedure
due to the overlapping of stellar point spread functions (PSFs). A
completeness test is carried out to correct for such an error and to
establish the actual magnitude distribution. We add artificial stars
to the mosaics and apply the same star extracting procedure to see
whether or not they can be recovered. Stars detected in each mosaic
are ranked based on their distances from neighboring stars. The
first 20 unsaturated stars with the largest separations are picked.
Each star's PSF is extracted from the mosaic and the local sky
background is subtracted. Finally, the individual PSFs are
re-sampled, shifted, and added to create a template PSF.

\begin{figure}  
\epsscale{0.5} \plotone{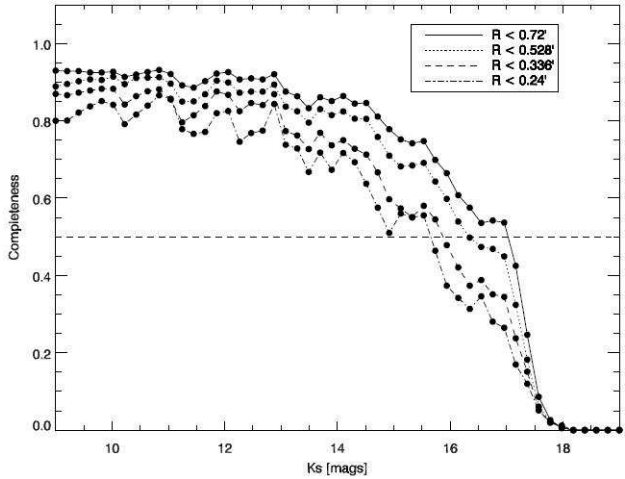} \caption{The completeness curves of
the derived photometry for the candidate cluster when choosing
different sizes. The curve for the outermost region shows that the
photometry is 50\% complete at Ks$\simeq$17. The dashed line
indicates 50\% completeness. \label{fig_completeness}}
\end{figure}

We divide the magnitude range from 9 to 19 magnitudes into bins with
a binsize of 1 magnitude. Within each bin, we create 50 fake stars
with their magnitudes uniformly distributed across the bin. The
images of the stars are created by scaling the template PSF and
added to the input mosaic at random locations within 43$''$ from the
center of the cluster field. The resulting mosaic is then treated
with DAOPHOT in the exactly same manner that is used to find stars
in the original extraction. If a star is found at the location where
a fake star is added and the difference between the input and output
magnitudes is less than 0.5 mag, we define that the fake star is
recovered. Otherwise, we define that the added artificial star is
missing. We repeat this experiment 10 times and generate 500
artificial stars total which are evenly distributed over the
magnitude bins. Figure~\ref{fig_completeness} shows the curve of
completeness from our simulation, which suggests that we achieve the
50\% completeness at M$_{Ks}{\simeq}$17~(mag) at the center of the
field.

\subsubsection{Spectroscopy}

We first subtracted dark frame from each science frame to minimize
fixed-pattern noise of the detector. Observations at two
complimentary nod-positions were then subtracted from each other to
remove atmospheric OH emission lines. Repeating observations were
averaged to reduce noise. Afterwards, spectra were extracted from
the average frame by summing the rows across the width of the
spectral traces. A spectrum of the arc lamp at the same position on
the chip was also extracted using the same extraction procedure.
Wavelength calibration was achieved by fitting the locations of the
arc lines with a 4th-degree polynomial. Atmospheric absorption in
the stellar spectra was removed using the telluric standard star,
which had first been divided through by a synthetic spectrum of a GV
star. Finally, stellar spectra were normalized by dividing through
by their median continuum values. The reduced spectra are shown in
Figure~\ref{fig_spec}. The coordinates of the stars are listed in
Table~\ref{tab_coor}. From flat regions of continuum, we estimate
the signal-to-noise of the spectra are 80 or better.

\subsection{Spitzer IRAC Photometry}

The basic calibrated data (bcd) was downloaded and used to create
new mosaics using Spitzer's Mopex tool before the mosaics were
treated with DAOPHOT. Point spread function (psf) photometry was
performed in order to achieve deeper photometric detection. After
every round of star finding process, the detected stars were
subtracted from the mosaic to create a residual mosaic. Additional
rounds of star finding were then performed on the resulting mosaics
until no additional stars were found in the final image. Finally,
all detected point sources were collected and cross-correlated with
2MASS point sources for completeness.

\section{Results}

\subsection{Color Magnitude Diagram}
\begin{figure}  
\epsscale{1.0} \plottwo{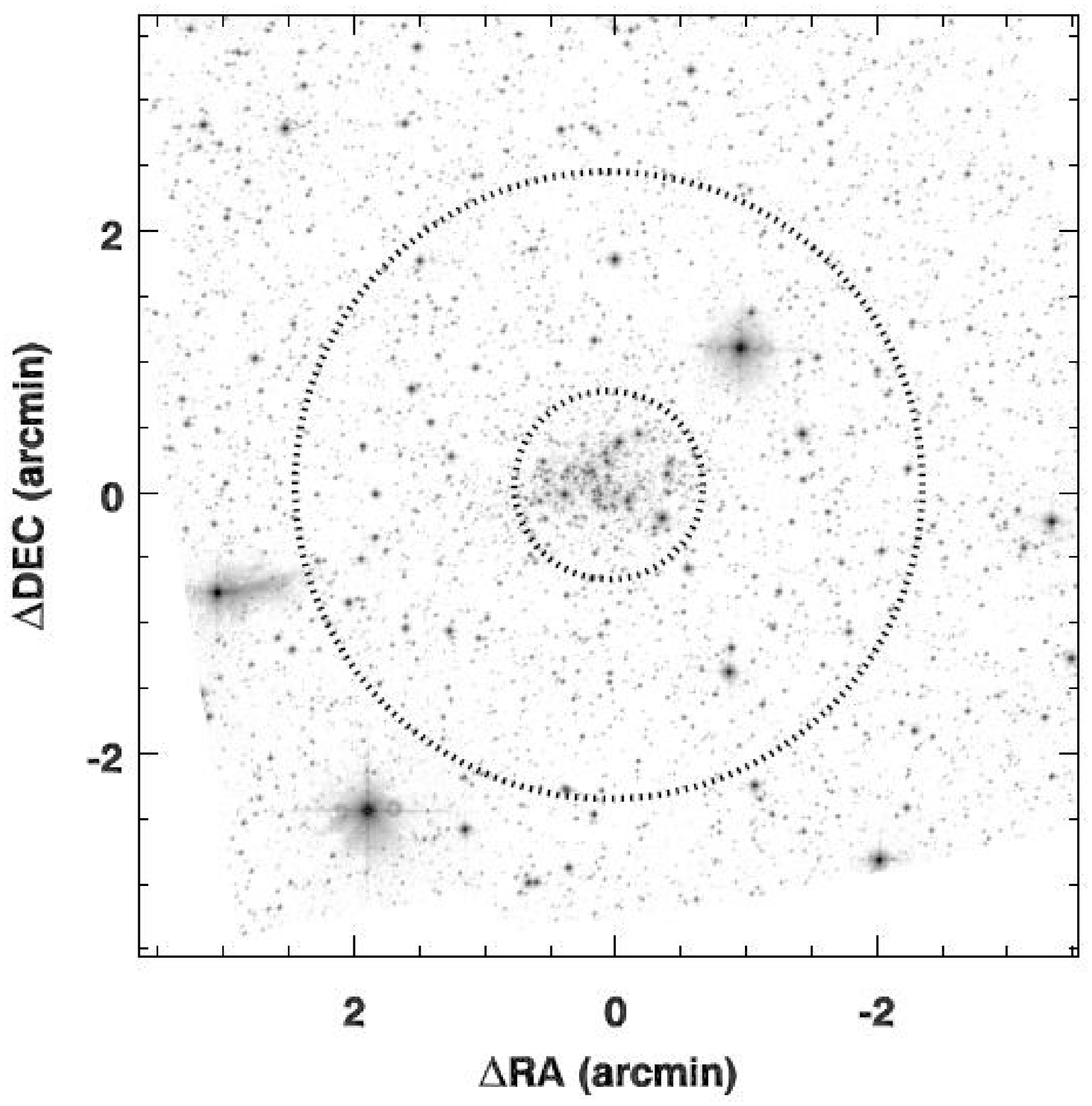}{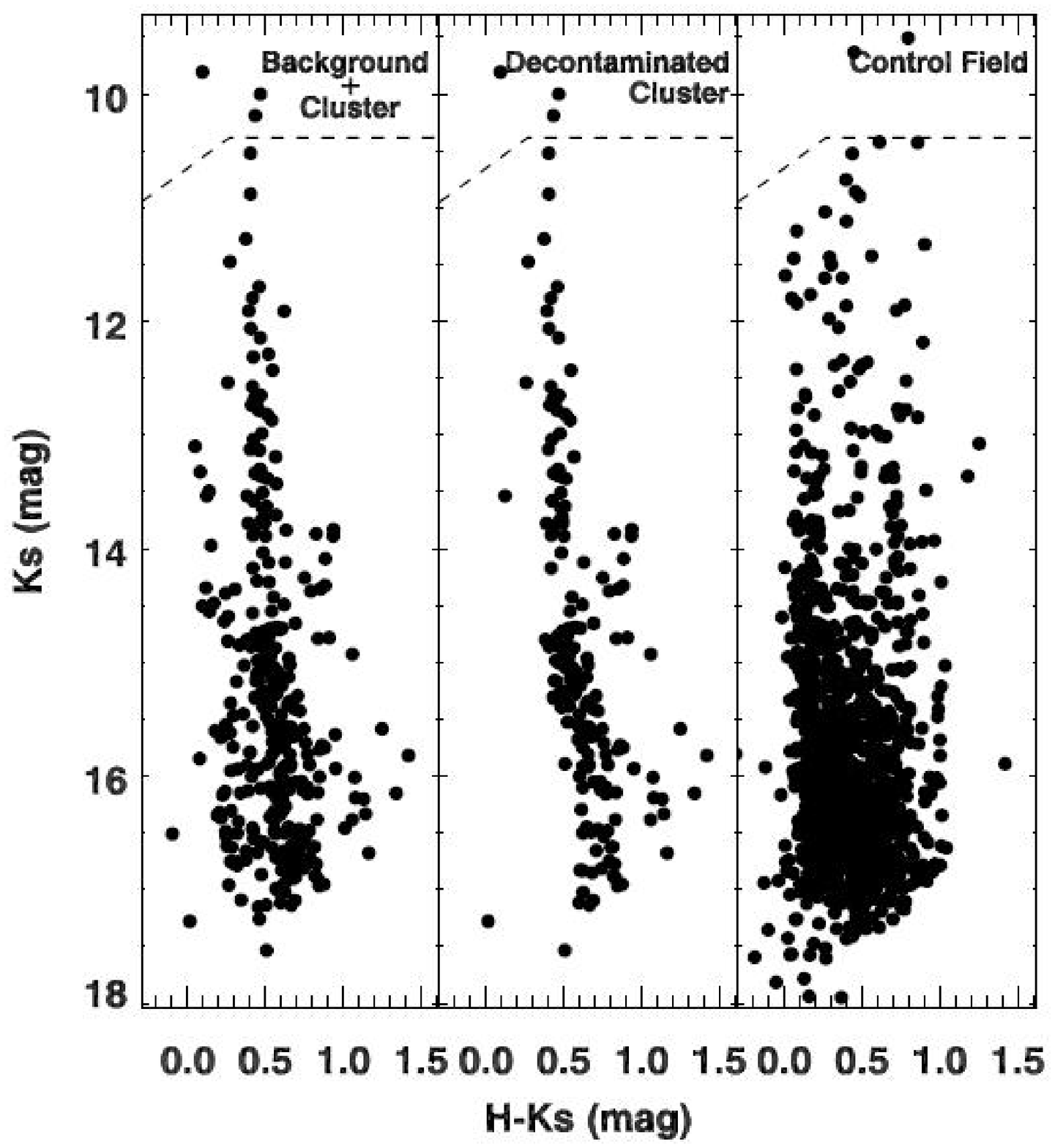} \caption{The
Color-Magnitude diagram (CMD) of [DBS2003]~45. The Ks band image (in
a logarithmic scale) on the left shows the region from which the
CMDs are drawn. Two concentric circles centered at (RA, DEC) =
(10$^h$19$^m$10.5$^s$, -58$^{\circ}$02$'$22.6$''$) demonstrate the
central circular region and an annulus region chosen for the cluster
candidate and the control field. The radii of the circles are 0.72
and 2.4 arcmin, respectively. The CMDs on the right in the figure
are for the cluster with background contamination (left panel), the
decontaminated cluster (middle panel) and the control field CMD
(right panel), respectively. \label{fig_cmd2}}
\end{figure}

Figure~\ref{fig_cmd2}(right) shows color magnitude diagrams (CMDs)
of the candidate cluster before (left panel) and after (middle
panel) field contamination has been removed and the CMD of a chosen
control field (right panel) for our SofI data. The image on the left
demonstrates how the regions for the cluster and the control field
are defined.
A circular region with a radius $\sim$0.72 arcmin is chosen for the
cluster and it includes the densest part of the observed field. An
annulus region outside the circular region is chosen as the control
field, which has an outer radius about 2.4 arcmin. Both regions are
centered at RA(2000)=10$^h$19$^m$10.5$^s$,
Dec(2000)=-58$^{\circ}$02$'$22.6$''$. The CMD of the control field
is scaled based on the area ratio of regions for the cluster and the
field and is subtracted from the CMD of the cluster+field. The
resulting decontaminated cluster CMD is shown in the middle panel of
Figure~\ref{fig_cmd2}(right), which demonstrates clearly a cluster
sequence with majority of stars along a vertical line with
H-Ks$\simeq$0.45. Among 175 remaining stars, the brightest one has
Ks$\simeq$10. Since a few stars are located above our saturation
limit (dashed lines in the CMDs), we have replaced presumably
saturated magnitudes with the corresponding values in the 2MASS
catalog, which does not significantly change the locations of those
stars on our CMDs.

\subsection{Spectroscopy}
\begin{figure}  
\epsscale{1.0} \plottwo{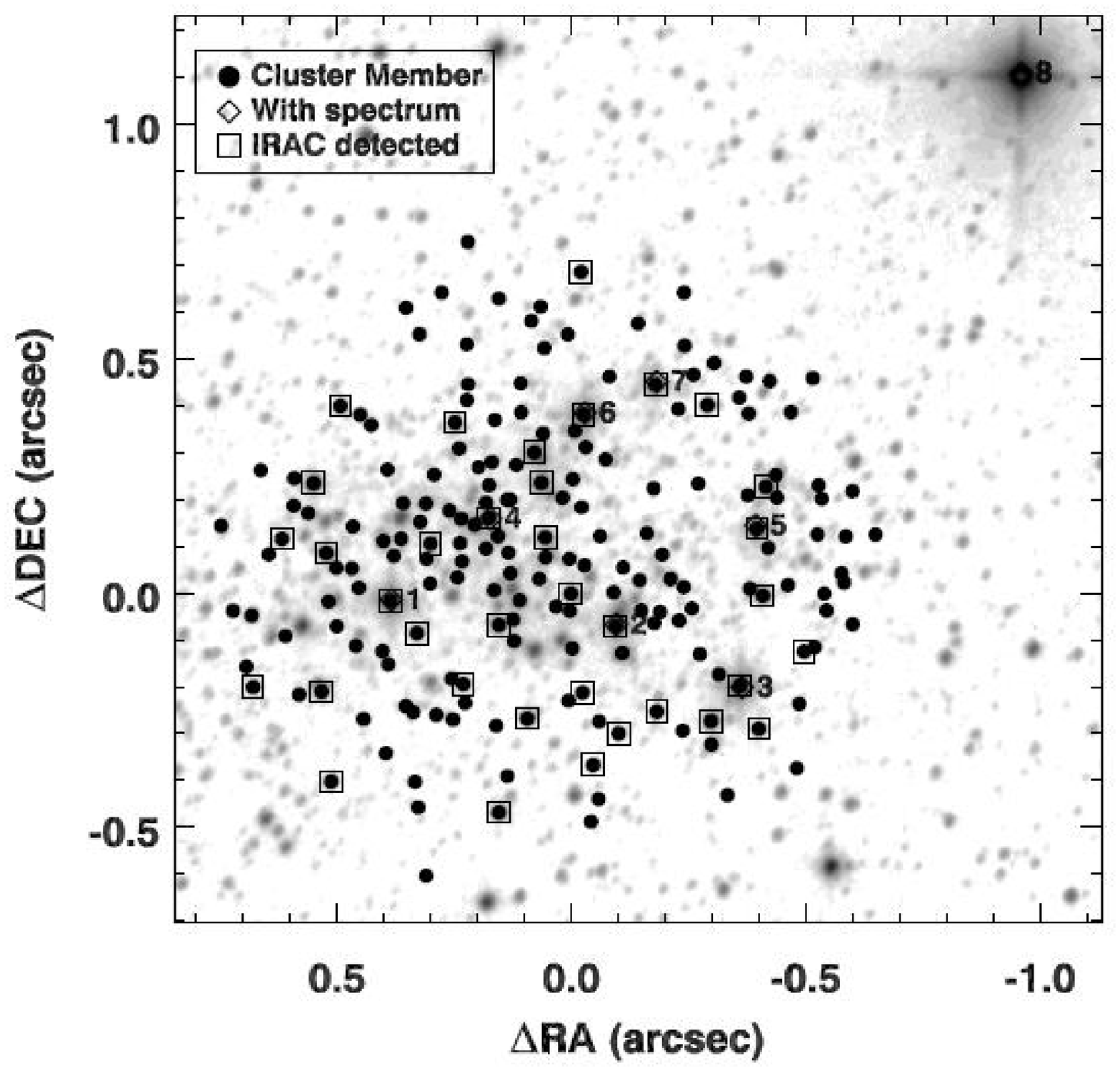}{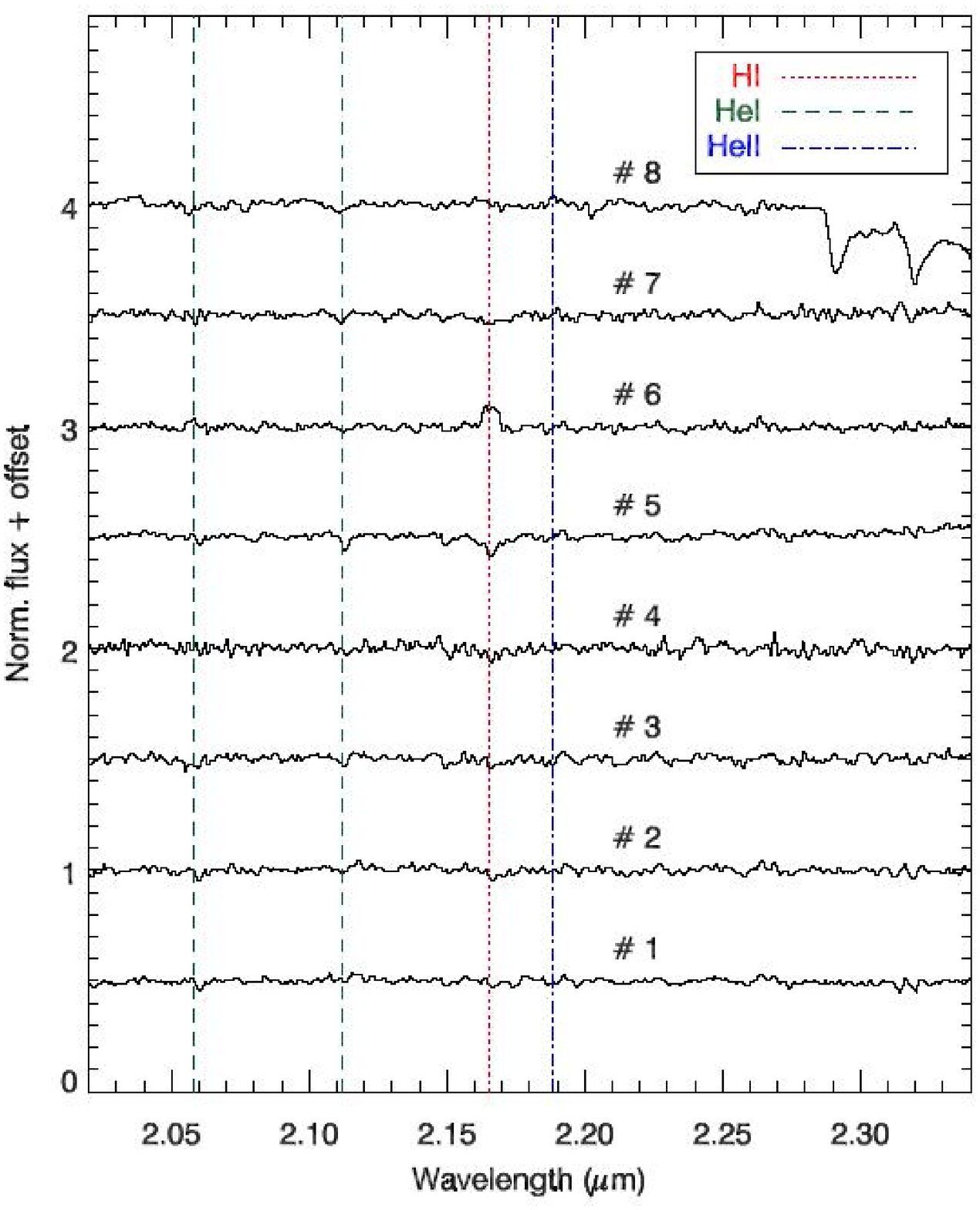} \caption{Low-resolution
spectra of a few picked stars. The image shows a blowup of the Ks
band mosaic of the cluster with symbols indicating different groups
of stars in the NIR CMD (Figure~\ref{fig_isochrone}) and the mid-IR
color-color diagram (Figure~\ref{fig_irac}) of the cluster. Filled
circles represent stars detected by SofI and stay in the CMD after
contamination subtraction. Squares represent stars detected by IRAC.
Diamonds and numbers indicate locations of stars with a spectrum
shown in the right panel. The image is centered at (RA, DEC) =
(10$^h$19$^m$10.5$^s$, -58$^{\circ}$02$'$22.6$''$).
\label{fig_spec}}
\end{figure}

Low-resolution spectra of a few relatively bright stars present in
the field of the candidate cluster are shown in the right panel of
Figure~\ref{fig_spec} with their locations indicated in a Ks band
image of the field on the left together with other stars remaining
in the cluster CMD. The spectrum of the brightest star in the field
\#8 shows CO bandhead absorption. The CO equivalent width is
consistent with the star being either a late-type giant (M5-7III) or
a supergiant of an earlier type (K2-4 I). The spectrum of the star
\#6 shows faint emission at the wavelengths of the Br$\gamma$ line
and the HeI 2.06~$\mu$m line and absorption in the HeI 2.11~$\mu$m
line. All other spectra show weak absorption at the wavelengths of
the Br$\gamma$ line, the 2.06~$\mu$m and 2.11~$\mu$m HeI lines. None
of these spectra shows significant contribution from the 2.19~$\mu$m
HeII line.

According to the atlas provided by \cite{hanKKPT05}, stars earlier
than O9.5 usually show HeII 2.19~$\mu$m line in absorption in their
spectra. This feature should be as strong as the Br$\gamma$ line
absorption in the early O type stars, fades away after O7 while
Br$\gamma$ gets stronger. The spectra of late O and early B type
stars should also possess the HeI 2.06~$\mu$m or 2.11~$\mu$m line in
absorption. These qualitative comparisons suggest the hot stars
shown in Figure~\ref{fig_spec} have spectral types later than O8,
probably early B types. Particularly, star \#6 has a spectral type
earlier than B0. The presence of the Br$\gamma$ line emission
indicates a spectral type around O9. Accurate spectral types need to
be determined by future high resolution spectroscopic observations
for these stars.

\subsection{Extinction}

The intrinsic H-Ks colors of most main sequence and evolved stars
are approximately zero. This allows us to estimate the average
foreground extinction from the observed H-Ks colors by assuming that
the observed H-Ks values of stars in a cluster sequence are entirely
due to the foreground extinction. In this way, we derive a
foreground visual extinction A$_V{\simeq}$7.1$\pm$0.5 assuming a
standard extinction law \citep{rieL85}.

Additionally, patchy extinction may be present at the observed field
because the distribution of stars is not symmetric. A hole of stars
can be seen at the west side of the identified cluster and more
stars seem to concentrate toward the southeast side of the cluster.

\subsection{Distance}
Without high resolution spectra, we have to rely on measurements in
the literature for the kinematic distance to the candidate cluster.
\cite{dehZC05} determined the kinematic distance of [DBS2003]~45 to
be $\sim$4.5~kpc from the Sun based on a radial velocity measurement
of a molecular cloud-HII region complex G283.9~-0.9, which seems
associated with the target based on their locations on sky
\citep{rus03}. This is the only distance measurement available.
\cite{bluCD00} gave the absolute K band magnitudes for main sequence
stars of different spectral types. At a distance of 4.5~kpc, late O
and early B dwarfs will have a M$_K$ in the range of 11 and 14 mags
with an assumed foreground extinction A$_V{\simeq}$7.1. The median
of these values is at least one magnitude fainter than our
measurement of $\sim$10.5~mags for those hot stars. This fact
suggests that the hot stars in our observations are giants or
supergiants, which is consistent with the presence of Br$\gamma$
emission.

The adopted distance puts the cluster candidate and the Sun at the
same radial distance from the Galactic center. Therefore, it is
reasonable to assume that the members of the cluster have similar
initial chemical abundances to solar values. For this reason we use
solar metallicity isochrones in the following analysis.

\subsection{Cluster Age}
In general, it is hard to determine the age of the cluster with only
NIR photometric data because of the degeneracy of isochrones at the
NIR wavelengths. However, the absence of early O stars and the
presence of several late O or early B type blue giants/supergiants
in our cluster candidate put some constraints on the cluster age.
Based on solar abundance Geneva evolutionary models \citep{lejS01},
we can determine that the cluster must be older than 3~Myr to allow
some massive stars to evolve off the main sequence. Our isochrone
fitting using solar abundance Geneva tracks suggests that for any
star earlier than O7.5 to be in our observed magnitude ranges, the
cluster needs to be younger than 6 Myr. We applied A$_V$=7.1 and a
distance of 4.5~kpc as suggested in the previous sections during the
isochrone fitting. Star \#8 in Figure~\ref{fig_spec} also provides
some constraint on the cluster age. Our spectroscopy suggests that
it is either a M5-7 giant or a K2-4 supergiant. The star is
saturated in our SofI photometry. The 2MASS photometry indicates
that the magnitudes of the star are H$\sim$6.57~Mag and
Ks$\simeq$5.85~Mag. If we take this red giant/supergiant as a
cluster member, the age of the cluster should be older than 8
Million years.

\begin{figure}  
\epsscale{0.5} \plotone{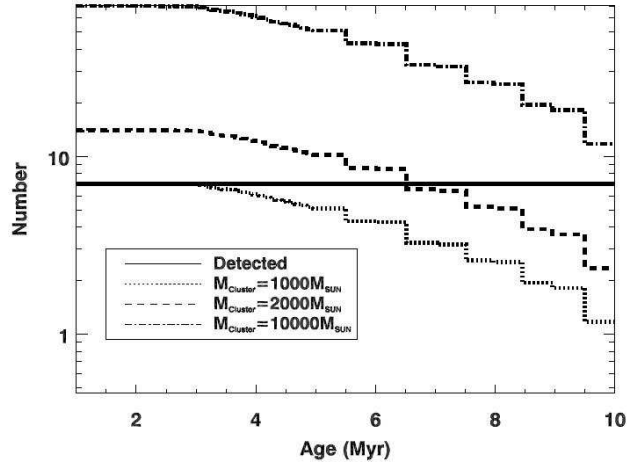} \caption{The number of stars with
masses over 15~M$_\odot$ versus the age in a cluster of mass 10$^3$,
2$\times$10$^3$ and 10$^4$~M$_\odot$, respectively. The `Salpeter'
IMF is assumed for the mass distribution at the zero age. The Geneva
isochrones with solar abundances are used for the evolution of
cluster members. The solid straight line indicates the number of
massive stars detected in [DBS2003]~45. \label{fig_sim}}
\end{figure}

To better constrain the age of the cluster, we carry out a
Monte-Carlo simulation to estimate the number of massive stars that
should be observed within our observing range (M$_K$$<$18 and
0.4$<$H-Ks$<$0.8) for a cluster. We create a stellar cluster which
has an IMF following the `Salpeter' law and has a total mass of
10$^6$~M$_\odot$. In order to improve number statistics, we use the
large cluster mass in the simulation. Afterwards, we scale the
resulting number down according to an assumed cluster mass. The
colors of stars in the cluster are computed for ages between 1~Myr
and 10~Myr according to the Geneva isochrones \citep{schSMM92}. The
same distance and average reddening as that derived for [DBS2003]~45
are assumed for the model cluster. Finally, we count the number of
stars with masses larger than 15~M$_\odot$, which is approximately
the initial mass of a B0.5 star \citep{bluCD00}. The results are
shown in Figure~\ref{fig_sim}. The curves are for a cluster of mass
10$^4$~M$_\odot$, 2.0$\times$10$^3$~M$_\odot$, and 10$^3$~M$_\odot$.
The simulation predicts that a cluster with
M$_{Cluster}\sim$10$^3$M$_\odot$ will have only 2-3 massive stars
present in our observing range when the cluster is older than 8~Myr.
This value is about two times smaller than our detection of 7-8
massive stars. This suggests that either the cluster mass is over
2000~M$_\odot$ or the cluster age is below 8~Myr. Using a 9.0~Myr
isochrone, we estimate that the mass of the cluster is around
1400~M$_\odot$. An alternative 6.0~Myr isochrone would suggest a
cluster mass of 1170~M$_\odot$. Therefore, the choice of the
isochrone does not cause a significant change in the resulting
cluster mass to be above 2000~M$_\odot$. We take this as the
evidence to support a younger age for the cluster.

From stellar evolution models, we know that all stars with a mass
M$<$120~M$_\odot$ of a 3~Myr old cluster should be still on the main
sequence. Therefore, there should be a limited number of giants or
supergiants in the cluster. For stars with an intermediate-mass
(M$\sim$25~M$_\odot$) to evolve off the main sequence, the cluster
should be at least 5~Myr old. Our detection of a few evolved stars
suggests that [DBS2003]~45 is older than 5~Myr. Therefore stellar
models support that [DBS2003]~45 is between 5 and 8~Myr old. This
analysis also suggests that the cool supergiant is not a member of
the cluster.

\begin{figure}  
\epsscale{0.5} \plotone{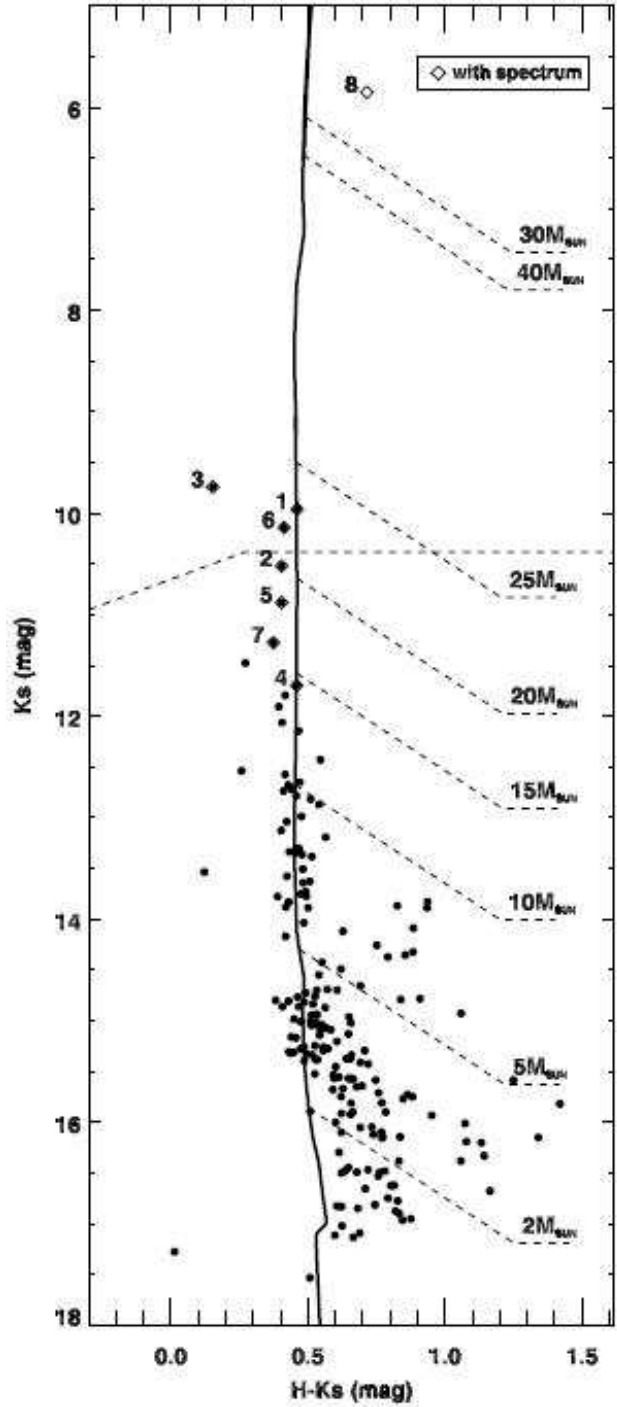} \caption{The CMD of [DBS2003]~45
with field contamination removed. The solid line shows the 6~Myr
isochrone. The stars with low resolution spectra are signified with
diamonds and numbers. The locations of stars of certain initial
masses are also indicated with the reddening vector.
\label{fig_isochrone}}
\end{figure}

In Figure~\ref{fig_isochrone}, we plot a 6.0~Myr isochrone on the
top of the CMD of the observed cluster. As in Figure~\ref{fig_cmd2}
and \ref{fig_spec}, the stars with available spectra are indicated
with diamonds and numbers. Open squares on the top of some stars
indicate that IRAC four color photometric measurements are also
available. In the figure, we also indicate the locations of a few
stars with different initial masses on the isochrone and the
direction of the reddening vector.

\subsection{Mass Function}
\begin{figure}  
\epsscale{1.0} \plottwo{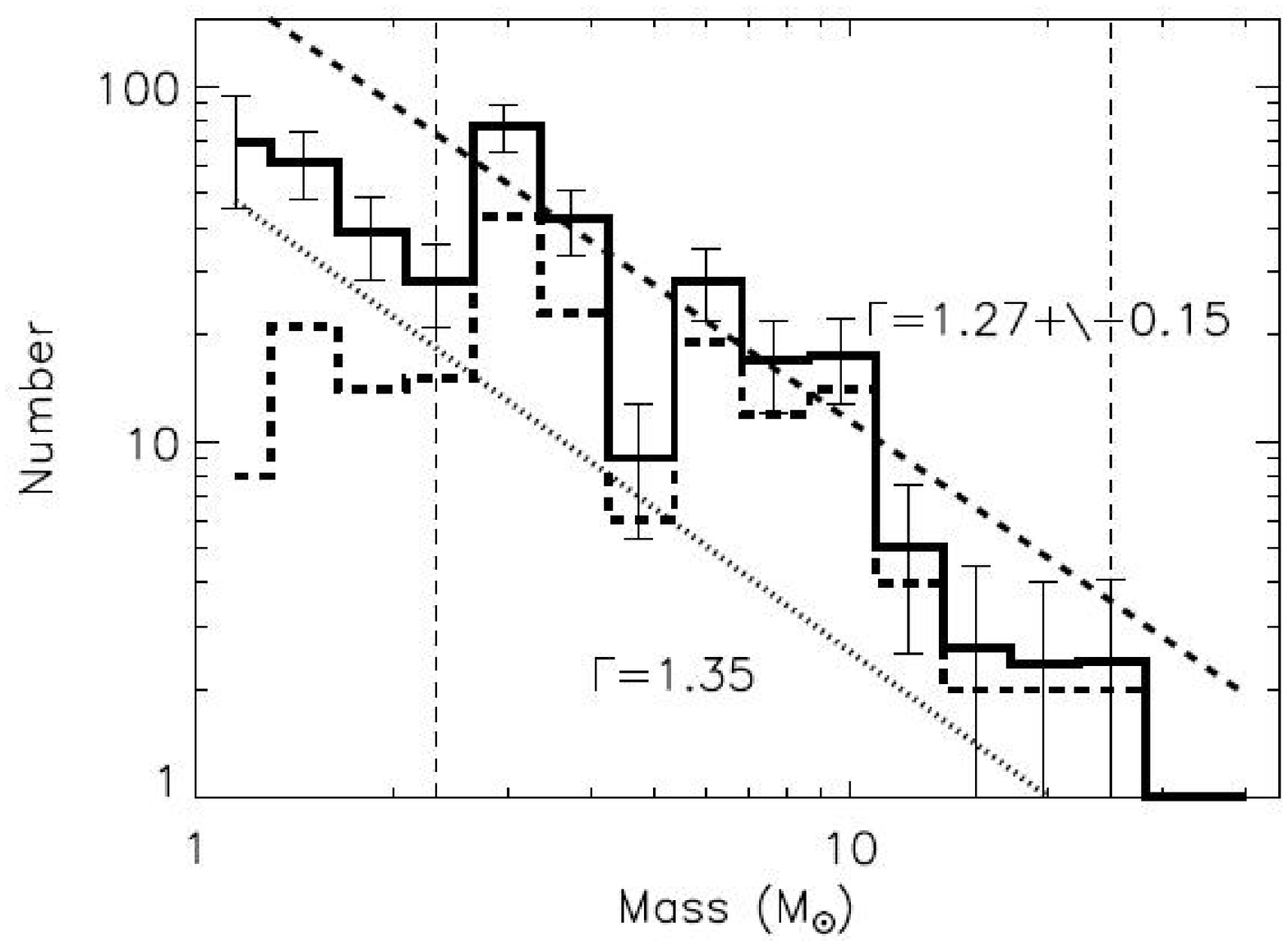}{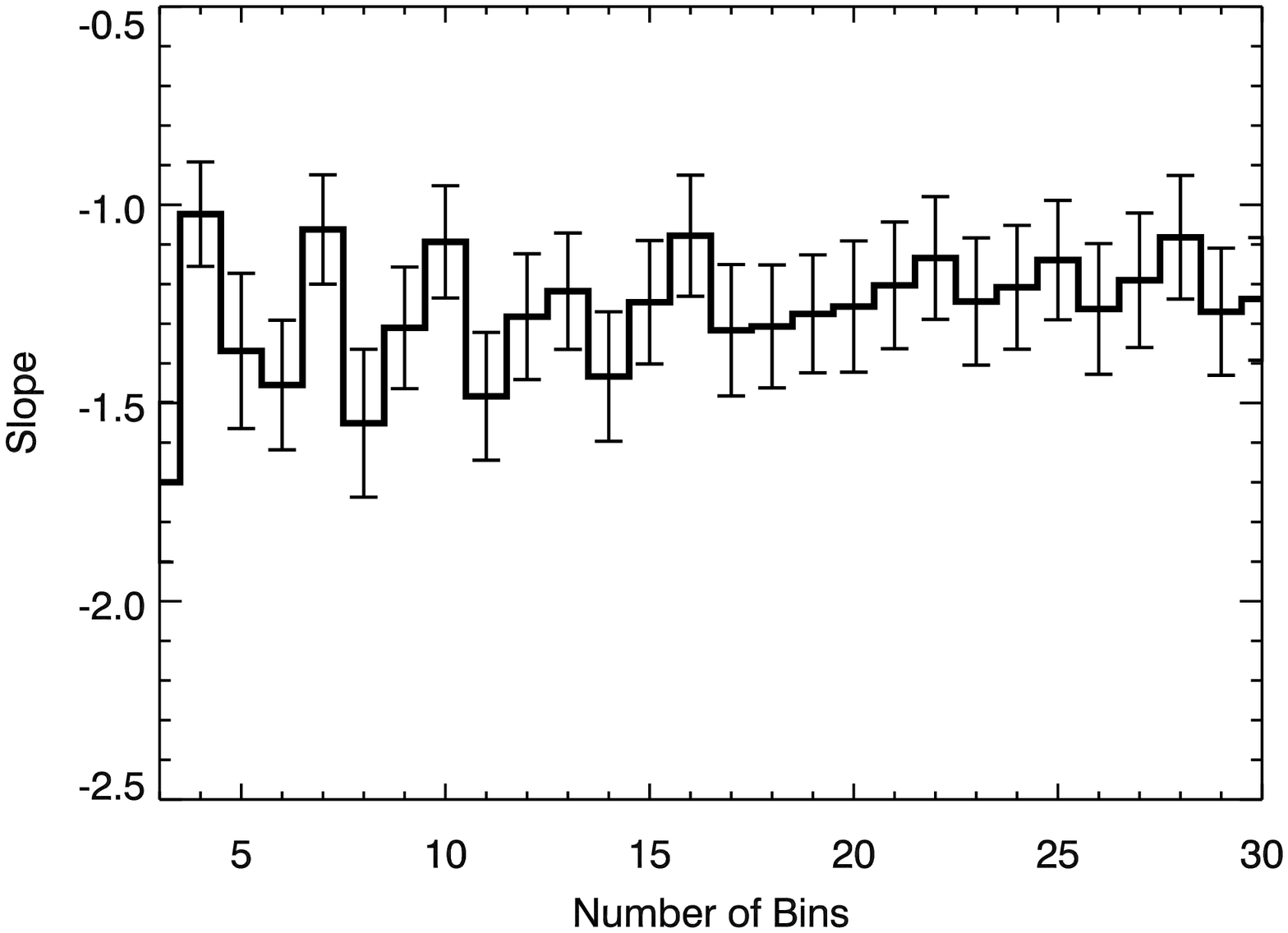} \caption{(Left) The mass
function of [DBS2003]~45 assuming a 6~Myr isochrone. The dashed
histogram shows the result before completeness correction and the
solid histogram shows the result after the correction. The fitting
result to a power law is indicated for the solid histogram. A
classic `Salpeter' model is also plotted. (Right) The slope of the
mass function with different bin numbers. \label{fig_mf}}
\end{figure}

The cluster mass function is based on the stellar magnitudes and the
chosen isochrones. The initial mass of a star can be immediately
determined if the star is close enough (within 5-$\sigma$
uncertainties) to the isochrone in the CMD. It is simply the
corresponding initial mass of the star with the same magnitude on
the isochrone. A few stars are on the left side of the isochrone.
They could be due to low number statistics. We leave them out in our
discussion. The initial masses of stars on the right side of the
isochrone are found by extrapolating the reddening vector back
toward the isochrone. Figure~\ref{fig_mf} (left) shows the mass
distribution for the final 164 stars with reasonable colors. The
Geneva 6~Myr isochrone with solar abundances is used to transform
colors to masses. The dashed line histogram shows the distribution
before the incompleteness correction and the solid line shows the
result after the correction. A set of mass bins equally spaced at
the logarithmic scale are chosen to make the histogram. A
theoretical (Salpeter) mass function of
$\frac{N}{N_0}=(\frac{M}{M_0})^{-\Gamma}$ with $N(20M_{\odot})=1$
and $\Gamma$=1.35 is also indicated with a straight line in the
figure. A regression procedure is used to fit the power law. The
regression only uses the bins with at least 3 counts and with a mass
bigger than the 50\% completeness limit. A slope of
$\Gamma$=1.27$\pm$0.15, which is close to the Salpeter value, is
found. We notice that the change of isochrone does not produce
significant difference in the slope. We estimate the total mass of
the cluster by taking the summation of the mass function between the
mass range from 0.5~M$_\odot$ to 45.0~M$_\odot$, which results in
$\sim$1092~M$_\odot$, $\sim$1170~M$_\odot$, and $\sim$1393~M$_\odot$
for the 3~Myr, 6~Myr, and 9~Myr isochrones, respectively.

Our field contamination removed CMD of [DBS2003]~45 has only $<$200
stars. The small size of the sample can result in large statistical
uncertainty, especially at the high mass end. There are only a
limited number of stars in the bins with the largest central masses.
We investigate the effect of the bin number on the resulting slope.
By choose different bin numbers, we divide the entire logarithmic
mass range into bins of equal sizes. We then populate the bins and
fit the resulting mass function to a power law. Figure~\ref{fig_mf}
(right) shows the result of this analysis. The horizontal axis is
the number of mass bins that are used. We notice that there is no
substantial change in the slope due to the change of the bin size
and the slope tends to stabilize at $\sim$-1.25 when the bin number
approaches 16 from both ends. Based on this analysis, we conclude
that the slope of the mass function is not very different from the
Salpeter value.

Compared to Arches cluster at the Galactic center, [DBS2003]~45 has
much lower mass. Therefore, our findings can not be used as the
evidence to support the statement that the mass function of massive
clusters is same as less massive clusters or field stars.
Nevertheless, the current data implies that intermediate or low mass
clusters may have a similar mass function as field stars. And, the
evolutionary effect such as evaporating low mass cluster members to
the field during the cluster lifetime seems not change the cluster
initial mass function significantly. Our result also does not
contradict with \cite{mor93}, who suggests that clusters near the
Galactic center may have shallower IMFs, since [DBS2003]~45 is at a
similar distance from the Galactic center as the sun.

However, our finding is based on only one cluster. Low number
statistics can be important in this case. Observations and analysis
of a large sample of clusters are needed to answer questions such as
whether or not the slope of the initial mass function is a function
of the cluster mass, the dynamical time and the environment.

\subsection{Mid IR Excess}

\begin{figure}  
\epsscale{1.0} \plottwo{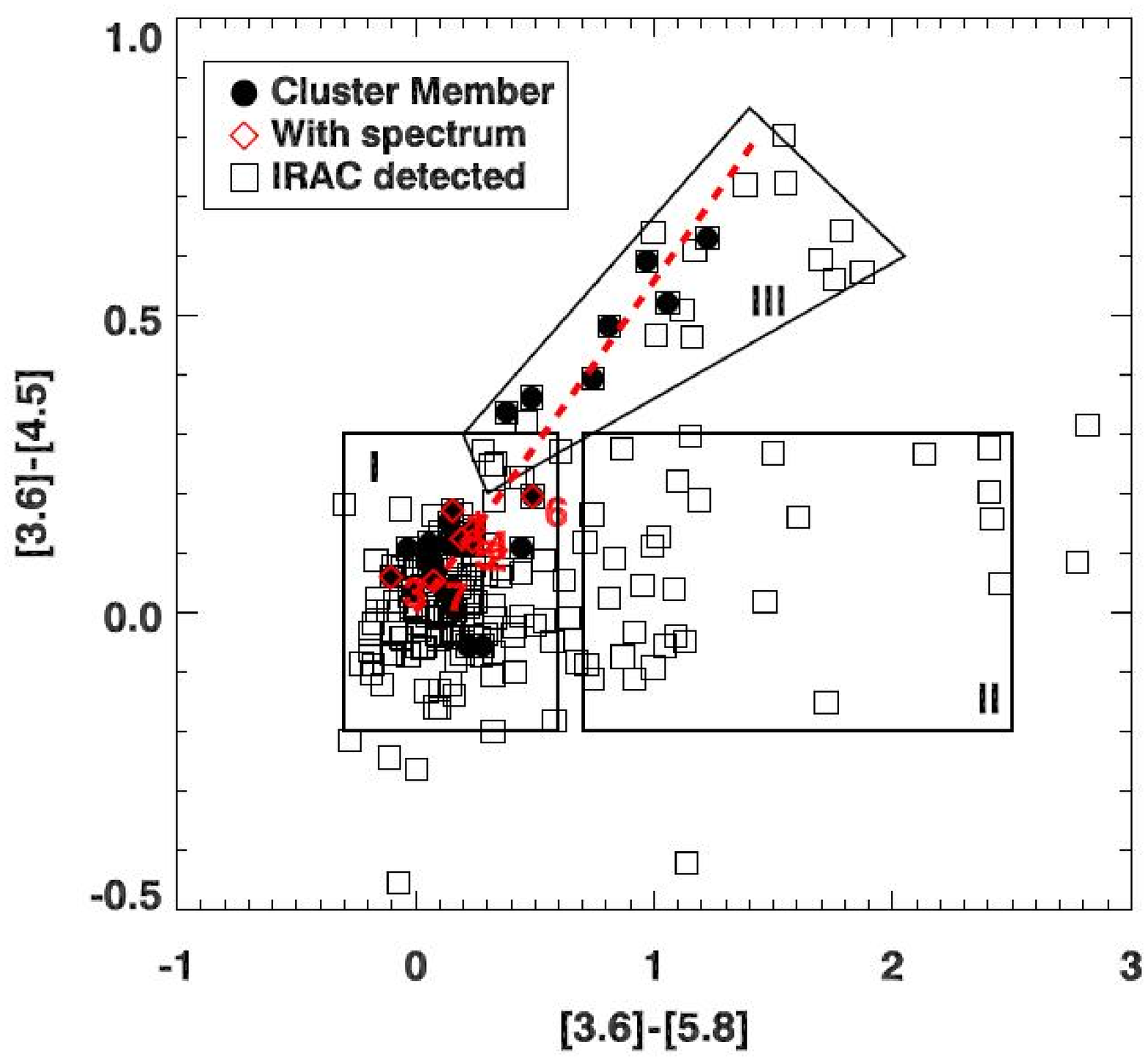}{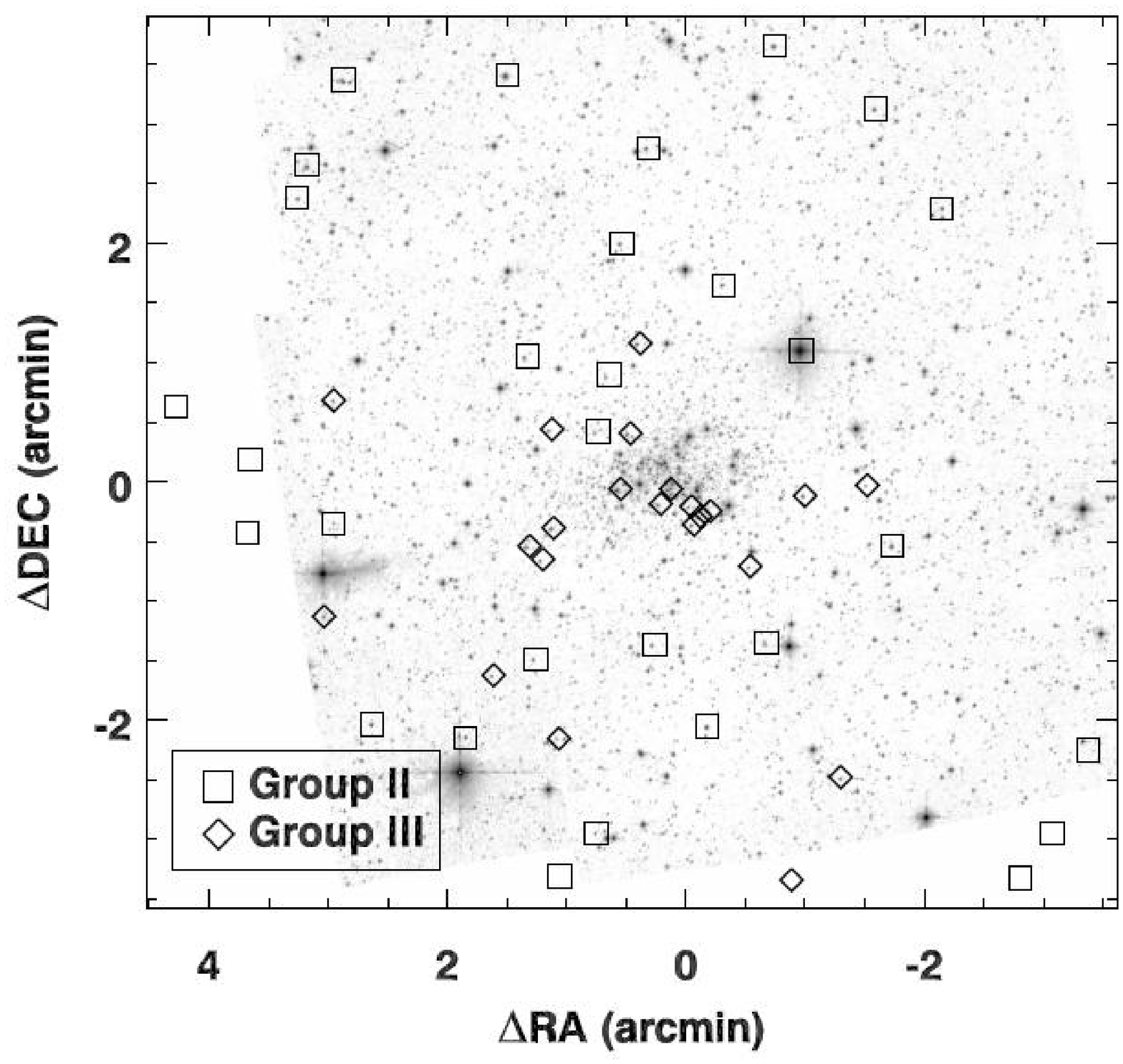} \caption{Left: The IRAC
color-color diagrams of the cluster. Squares represent data points
in the IRAC photometry and filled circles represent the stars with
SofI measurements. Stars with spectra are indicated with diamonds
and numbers. Right: The locations of stars showing 5.8~$\mu$m excess
(Group {\bf II}) and strong foreground dust extinction (Group {\bf
III}) are indicated with squares and diamonds, respectively. The
image is centered at (RA, DEC) = (10$^h$19$^m$10.5$^s$,
-58$^{\circ}$02$'$22.6$''$). \label{fig_irac}}
\end{figure}

As we mentioned earlier, a certain degree of asymmetry in spatial
distribution is observed for cluster members. This could suggest
patchy extinction toward the field. To confirm this, we plot the
photometry of stars obtained by Spitzer/IRAC in
Figure~\ref{fig_irac}. We can divide stars on the IRAC M$_{3.6-4.5}$
versus M$_{3.6-5.8}$ color-color image into three groups, which are
indicated by solid-lined boxes. Majority of stars in the lower-left
box have M$_{3.6-4.5}$$\sim$0.0 and M$_{3.6-5.8}$$\sim$0.0 (group
{\bf I}). In the second box, stars have almost constant
M$_{3.6-4.5}$ but varying M$_{3.6-5.8}$ (group {\bf II}). In the
third box, M$_{3.6-4.5}$ seems to vary linearly with M$_{3.6-5.8}$
and with $\frac{M_{3.6-4.5}}{M_{3.6-4.5}}{\sim}$0.3 (group {\bf
III}).

Stars in group~{\bf I} include almost all stars on the cluster
sequence, particularly those close to the assumed isochrone. It is
consistent with their small color differences. In the figure, we
plot the reddening vector with A$_{Ks}$=6.0 mags assuming the
intrinsic M$_{3.6-4.5, 0}$=M$_{3.6-5.8, 0}$=0.0 and a standard
extinction law (A$_{\lambda}{\varpropto}{\lambda}^{-1.53}$). With
A$_{Ks}{\sim}$0.112A$_{V}{\sim}$0.8 mag, the values of M$_{3.6-4.5}$
and M$_{3.6-5.8}$ would be small and within the box we plotted.
Group {\bf II} represents stars with 5.8~$\mu$m excess. The
5.8~$\mu$m and 8.0~$\mu$m channels of IRAC were designed to observe
emission from Polycyclic Aromatic Hydrocarbon (PAH) particles in
warm interstellar medium. The detection of 5.8~$\mu$m excess may
suggest a group of stars with strong PAH emission from their
circumstellar envelopes. One interesting fact about these stars is
that none of them is present in our cluster CMD. We plot their
locations on the top of our Ks band image of the field to check the
integrity of our results (right panel of Figure~\ref{fig_irac}),
which suggests that those 5.8~$\mu$m bright stars (squares) are all
at large distances from the cluster center, including some areas
which are not covered by our NIR observations. Thus, it is not a
surprise that they are excluded from the cluster population. These
stars could be red giant stars in the field, which have lower than
main sequence stellar effective temperatures. Stars in group {\bf
III} distribute approximately along the reddening vector and seems
to suggest that their colors are strongly affected by varying
foreground extinction. Their locations in the Ks band map indicate
they do distribute unevenly in the cluster, supporting a patchy
extinction interpretation.

Stars left unexplained are those faint in the Ks band and heavily
reddened in Figure~\ref{fig_isochrone}. None of them have an IRAC
correspondent. They are probably too faint for IRAC to pick up since
the detection limit of IRAC photometry is Ks$\simeq$15.3 mags based
on our cross-correlation result between the IRAC catalog and the
SofI catalog. Their uneven spatial distribution in the cluster
demonstrated by Figure~\ref{fig_spec} is similar to that of stars in
group {\bf III} in the Figure~\ref{fig_irac}. Therefore, the extra
red colors of these faint stars may also be caused by foreground
extinction, although we could not rule out the possibility that they
have intrinsic colors similar to those in group {\bf II}. Deeper
photometry is needed to clarify on this.

\section{Discussion}

\subsection{The Cluster's Ionized Bubble}
Ionized gas has been observed around massive clusters. In some
cases, well-defined bubbles of ionized material and warm dust are
seen in radio and infrared wavelengths \citep{dehZC05}. It is
suggested that dusty material around clusters is heated by energetic
photons from cluster members to high temperatures and emits at
infrared wavelengths. By examining the properties of radio/IR
continuum from the heated material, one can check the consistency
between the total energetics and the assumption about the cluster
members.

\begin{figure}  
\epsscale{1.0} \plottwo{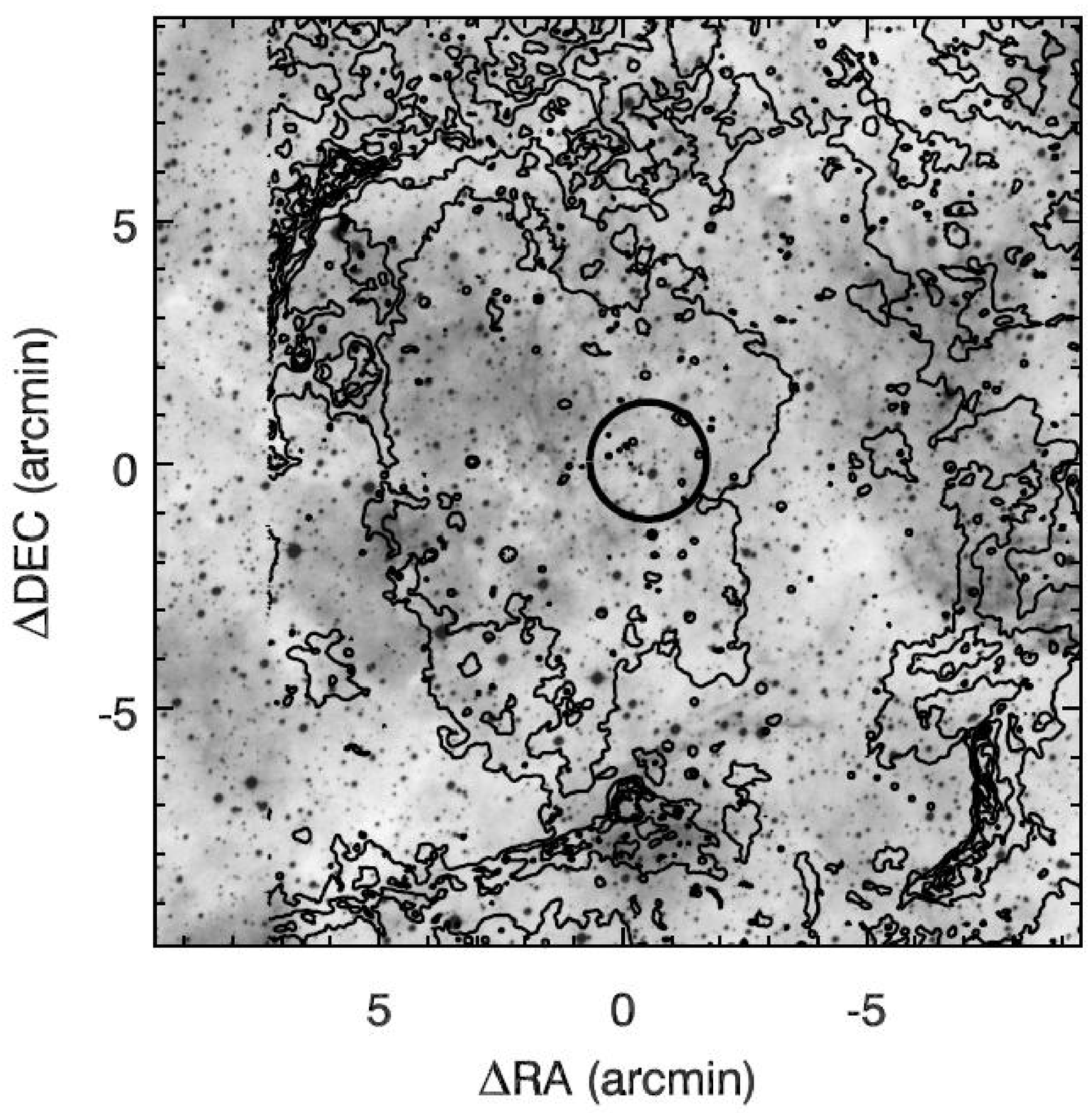}{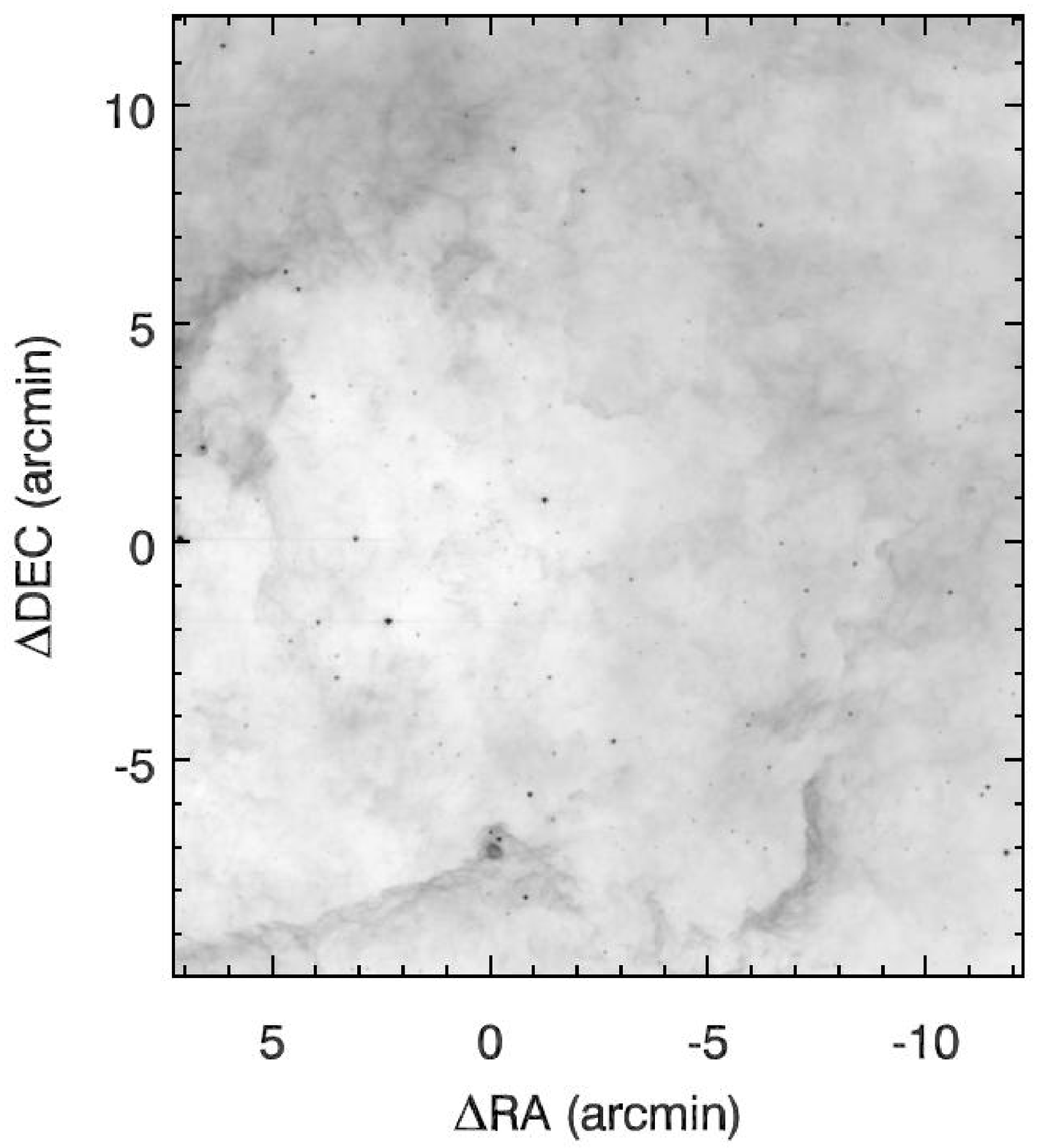} \caption{Left: The
contours of the IRAC 8.0~$\mu$m image are plotted over the
SUPERCOSMOS H$_\alpha$ grey scale image. It is evident that emission
from neutral material is mainly come from a ring-like structure
around the concentration of emission of ionized gas identified by
H$_\alpha$ line emission. The location of the cluster is indicated
with a circle. Right: The IRAC 8.0~$\mu$m image is shown in grey
scale. Point sources can be seen at several locations along the rim
of emission. Both images are centered at (RA, DEC) =
(10$^h$19$^m$10.5$^s$, -58$^{\circ}$02$'$22.6$''$).
\label{fig_bubble}}
\end{figure}

We over-plot Spitzer IRAC 8.0~$\mu$m emission contours on the top of
superCOSMOS H$_\alpha$ emission grey scale image
(Figure~\ref{fig_bubble}, left). A well-defined cavity of 8.0~$\mu$m
emission can be seen around relatively concentrated H$_\alpha$
emission. The location of the identified cluster inside the bubble
suggests that the cluster is the source of the ionizing photons.
Radio VLA 4850~MHz continuum observations showed that the total
continuum flux density of the HII region is 4519$\pm$99~mJy
\citep{wriGBE94}. In a thermal situation, radio continuum flux
luminosity (L$_{\nu}^{ff}$) and flux density (F$_{\nu}^{ff}$) are
related to the total Lyman continuum photon flux, Q$_{Ly}$
\citep{rybL86}:
\begin{eqnarray}
L_{\nu}^{ff}&=&4{\pi}D^2F_{\nu}^{ff}=6.8{\times}10^{-38}T^{0.5}e^{-h{\nu}/kT}\bar{g}_{ff}{\int}N_eN_idV
\\
Q_{Ly}&=&{\alpha}_B{\int}N_eN_idV.
\end{eqnarray}
Here D is the heliocentric distance to the HII region,
T$\simeq$10$^4$K electron temperature, $\bar{g}_{ff}$=5.1 the Gaunt
factor, N$_e$ and $N_i$ the electron density and the proton density,
and ${\alpha}_B$ the recombination coefficient. The integral is over
the entire volume of the ionized region. We can derive the ionizing
photon luminosity to be
Q$_{Ly}{\simeq}$8.2$\times$10$^{48}$~sec$^{-1}$. This luminosity
corresponds to the total ionizing luminosity of one O8.5 giant star
or four B0 type giants \citep{schD97,steHP03}. This is consistent
with our detection of $\sim$6 early B giants plus one O9 giant. The
total energy budget is also consistent with the value from far-IR
observations. \cite{dehZC05} reported that the far-IR fluxes of IRAS
point source associated with the dust ring indicates that the total
energy required to power the IRAS source is equivalent to the total
luminosity of 10 B1.5V stars. Nevertheless, this energy budget
should be the lower limit because it is possible that some fraction
of the ionizing photons would escape from the envelope of the
surrounding neutral material. From our analysis, it is plausible
that the HII region and the bright dust rim around [DBS2003]~45 are
excited by the central cluster.

\subsection{Triggered Star Formation}
Observations indicate that the spatially separated subgroups in
given OB associations show systematic differences in stellar age
\citep{bla64}. \cite{elmL77} suggest that the formation of these
subgroups is triggered sequentially by the expansion of HII regions.
Other scenarios of triggered star formation on different scales have
also been proposed \citep[][and references therein]{elm98}. As we
know, in spontaneous star formation the fragmentation and collapse
of molecular clouds is due to the loss of internal hydrostatic
equilibrium between gravity and supporting forces. On the other
hand, external dynamical processes dominate the compression of
molecular clouds and initiate the following fragmentation and
collapse in triggered star formation scenarios. Triggered star
formation is believed to be ubiquitous across star forming regions
and galactic disks \citep{elm98}.

Among the proposed scenarios, the ``collect and collapse'' process
at the peripheries of HII regions is particularly interesting
because it is thought to be a mechanism by which very massive stars
can form \citep[][, hereafter W94]{whiBCDT94}. In this scenario,
star formation is triggered by the propagation of an
ionization-shock front into molecular clouds. As the shock front
drives into the clouds, neutral material is accumulated and
compressed in a layer behind the front. Eventually, this layer of
gas becomes gravitationally unstable, fragments, and collapses to
form stars. Relatively higher temperature of the shocked gas causes
the collapse to start at a higher Jean's mass.

Despite its appealing to the theory of massive star formation, the
collect and collapse process is not convincingly confirmed due to
the lack of evidence. This is partially because young stellar
objects in the peripheries of HII regions are usually embedded in
molecular material and can not be easily detected at optical
wavelengths. It is also difficult to distinguish it from other
triggering mechanisms since they all introduce systematic
differences of the physical properties, e.g. stellar age and mass,
between the formed subgroups. With appropriate initial conditions,
even spontaneous star formation can cause similar observational
features. The initial hierarchical structure in large star forming
regions can result in a systematic progression in size, age, and
velocity dispersion for OB subgroups formed through the spontaneous
process \citep{elm98}. Particularly, the collect and collapse
process and the external pressure induced collapse of pre-existing
clumps occur on the same scale and cause radial progressions in
physical properties, therefore are difficult to be distinguished
from each other.

\cite{elm98} points out that the collect and collapse process
differs from the clump-squeezing process in two aspects. In the
former scenario, star formation activities are more like a burst
along the ridge of the accumulated gas. In contrast, star formation
is continuous throughout the disturbed cloud in the latter case. The
collect and collapse scenario tends to form stellar clusters and no
star at an intermediate age should exist between the clusters, while
the clump compression mechanism tends to form individual stars or
small stellar systems and the age distribution of stars is smooth.
Therefore, it is possible to distinguish two processes using these
properties. We should note that different mechanisms of star
formation may be present at the same time. The shock front of an HII
region can compress pre-existing globules inside molecular clouds
and force them to collapse. At the same time, ambient medium between
the globules is swept-up into a layer which can form stars through
the collect and collapse process. Nevertheless, observations of
young stars along the ridges of HII regions provide supporting
evidence of star formation through the collect and collapse
mechanism since stars formed through other mechanisms will not
distribute preferentially along the shock front.

In a search for infrared sources at HII region peripheries,
\cite{dehZC05} observed bright and red objects in the dust rim
surrounding the ionized gas around [DBS2003]~45. These authors
suggest that the point sources are the results of the collect and
collapse star formation. Our new observations of [DBS2003]~45 with
Spitzer IRAC show a ring-like structure in all four wavelength
channels surrounding the cluster. Many bright sources can be seen
along the ridge of the structure (Figure~\ref{fig_bubble}, right).
The light in these mid-IR channels is generally attributed to large
molecules such as polycyclic aromatic hydrocarbons (PAHs)
\citep{legP84}. The observed morphology clearly indicates that the
ring-like structure is the boundary of the ionized gas and very
likely consists of neutral material accumulated during the expansion
of the ionization front. Bright and red stellar objects spotted
along this structure are the best example of triggered star
formation in the swept-up layer around HII regions. Therefore, our
observations support that the collect and collapse star formation
process is undergoing in [DBS2003]~45. Further support for this star
formation mechanism in action would come from observations of the
age and mass distributions of the stars along the HII region
peripheries. These stars should be significantly younger than the
stars belonging to or near the cluster. Observations of similar
masses and separations for the newly formed stars along the
peripheries would support that they are formed together through the
fragmentation of the compressed layer since this process occurs on a
typical length scale and produces molecular cores with a typical
mass \citep{whiBCDT94, elm98}.

\subsection{Comparison With A Theoretical Model}

\cite{whiBCDT94} studied the fragmentation of the dense layer of
accumulated material surrounding an expanding HII region and derived
formula for the time at which the fragmentation starts (t$_{frag}$),
the radius of the HII region at that time (R$_{frag}$), the column
density of the layer, the masses of the fragments (M$_{frag}$) and
the separation between them. Particularly, W94 showed that both
t$_{frag}$ and R$_{frag}$ depend weakly on the isothermal sound
speed (a$_S$) in the compressed layer, the number of Lyman continuum
photons emitted by the exciting star per second (Q$_{Ly}$) and the
density of the ambient medium (n$_0$):
\begin{eqnarray}
\label{eq_rfrag}
R_{frag}&{\simeq}&5.8~a_{.2}^{4/11}~Q_{49}^{1/11}~n_3^{-6/11}~pc\\
\label{eq_tfrag}
t_{frag}&{\simeq}&1.56~a_{.2}^{7/11}~Q_{49}^{-1/11}~n_3^{-5/11}~Myr,
\end{eqnarray}
where a$_{.2}$=[a$_S$/0.2~km s$^{-1}$], Q$_{49}$=[Q$_{Ly}$/10$^{49}$
s$^{-1}$] and n$_3$=[n$_0$/10$^{3}$ cm$^{-2}$] are dimensionless
variables. They also predicted that the mass of fragments depends
strongly on the sound speed a$_S$:
\begin{eqnarray}
M_{frag}&{\simeq}&23~a_{.2}^{40/11}~Q_{49}^{-1/11}~n_3^{-5/11}~M_{\odot}.
\end{eqnarray}
This model has been tested by \cite{dalBW07} using an 1-D smooth
particle hydrodynamical code. Although M$_{frag}$ predicted by
\citeauthor{dalBW07} is $\sim$2.5 times smaller than the value
predicted by the W94 model, the predicted t$_{frag}$ and R$_{frag}$
from two models agree with each other within 20 to 25 per cent,
suggesting that the W94 model is valid and accurate.

\cite{zavDCB06}(hereafter Z06) applied the analytical model to a
real HII region RCW~79 to estimate its age and other properties.
Located at 4.2~kpc \citep{rus03}, RCW~79 appears to be a circular
HII region surrounded by a partially open dust ring. Point sources
are observed along the dust ring. This morphology of RCW~79 suggest
that the expansion of the HII region triggered the star formation in
the dust ring as described by the collect and collapse process
\citep{dehZC05}. Z06 compared the 1.2~mm continuum, NIR and mid-IR
data of RCW~79 with the analytical model and estimated that the
dynamical age of RCW~79 is 1.7~Myr. They also estimated that the
triggered star formation along the dust ring occurred $\sim$10$^5$
years ago. These timescales and the corresponding density
($\sim$2000~cm$^{-3}$) of the surrounding molecular material are
used to constrain the age of an associated compact HII region and
the masses of fragments observed in the ring. The good agreement
between the observations and the model predictions indicates that
the collect and collapse process is in deed the main triggering
mechanism of star formation in the dust ring around RCW~79.

We have not performed molecular line observations to help us
constrain the masses of fragments and the sound speed inside the
molecular layer in [DBS2003]~45. However, using the morphology of
[DBS2003]~45 and the theoretical predictions of W94 (similar to the
study of RCW79 by Z06), we can compare the fragmentation age of the
expanding bubble with both the bubble's dynamical age and the age
derived for the cluster. The Spitzer IRAC images of [DBS2003]~45
show that the average radius of the ring-like structure is about
8~arcmin. This corresponds to a fragmentation radius of
$\leq$10.5~pc at a distance of 4.5~kpc. Adopting an intermediate
value (0.4~km~s$^{-1}$) for a$_S$, we can derive
n$_0\simeq$500~cm$^{-3}$ and t$_{frag}\simeq$3.3~Myr, based on
Eq.~\ref{eq_rfrag} and \ref{eq_tfrag}. Since we assume that the
molecular cloud is uniform with a small sound speed and the
expansion of the HII region is continuous, the derived n$_0$ should
be considered as the lower limit. Here we have applied
Q$_{Ly}$=8.2$\times$10$^{48}$~photons per second. Additionally,
according to W94 and \cite{dysW97}, the radius of the swept-up shell
around an evolved HII region grows with time $t$:
\begin{eqnarray}
\label{eq_rhii}
R{\simeq}4.5~Q_{49}^{1/7}~n_3^{-2/7}~t_M^{4/7}~pc,
\end{eqnarray}
where t$_M$=[t/Myr]. Adopting R=10.5~pc, Q$_{49}$=0.82 and
n$_3$=0.5, we have $t\simeq$3.3~Myr. The above age estimates are
consistent with each other and suggest that the point sources along
the dust ridge formed not long time ago. A typical sound speed in
the compressed layer is around 0.2-0.6~km~s$^{-1}$. We used
a$_S$=0.4~km~s$^{-1}$ in the above calculation, which implies an
uncertainty of $\pm$1~Myr for the derived fragmentation age. We have
also assumed other values, e.g. the density of the compressed layer,
the ionizing luminosity of the central cluster and the size of the
ionized bubble, which may introduce additional uncertainty into the
derived dynamical age and the fragmentation time of the bubble. If
we take the uncertainty into account, 3.3~Myr is in a good agreement
with the age of the central cluster of 5~Myr that we have derived in
the earlier section.

The above derived numbers rely on the following assumptions: the
ionizing photon luminosity of the exciting source is constant during
the lifetime of the HII region; the HII region expands within an
originally uniform and smooth molecular cloud; the compressed layer
by the ionization/shock front has a typical sound speed in dense
molecular clouds; the effects of stellar winds and other more
complex physical processes are not considered. These assumptions may
be oversimplified compared to real star forming regions. More
rigorous calculation of the nebular dynamics is well beyond the
scope of the current work. Nevertheless, our simple calculations
suggest that the radio emission from the HII region is consistent
with being powered by the ionizing flux from the cluster. The
dynamical age and the fragmentation age of the expanding bubble are
consistent with the age we derived for the central cluster. These
findings together point to a scenario in which the central cluster
has driven the expansion of the HII region into the surrounding
medium, and a second generation of star-formation is currently being
triggered through the collect and collapse mechanism at the
periphery of the bubble.

\section{Summary}

We present a photometric and spectroscopic study of a stellar
cluster candidate on the southern sky. Our study confirms that
[DBS2003]~45 is a cluster of a total mass $\sim$10$^3$~M$_\odot$.
The field decontaminated CMD clearly shows a cluster sequence at
H-Ks$\simeq$0.45, suggesting an average extinction Av$\sim$7.1~Mags.
Based on low-resolution spectroscopy, we identify a few bright
cluster members to be early type blue giants/supergiants. The
presence of these late O / early B type stars and the absence of any
early O type stars and M-type supergiants, suggests an age between 5
and 8 Myr for the identified cluster. A mass function with an
approximately Salpeter slope is found for the cluster. The
associations of the cluster with the SuperCOSMOS H$_\alpha$ emission
morphology and the ring-shaped 8.0$\mu$m structure observed by the
infrared camera IRAC on the Spitzer telescope support that the
cluster is relatively young and the ambient medium has not been
completely dispersed by stellar winds and supernova explosions.
Evidence of patchy extinction toward the cluster is found and is
consistent with the presence of a population of heavily reddened
stars. The presence of infrared point sources along the rim of the
dusty bubble supports the mechanism of triggered star formation.
Simple calculations of the radio emission and the ages of the HII
region support that the identified cluster is the source to drive
the expansion of the HII region into the ambient medium and to
trigger the formation of new stars.

\acknowledgments The material in this paper is supported by NASA
under award NNG 05-GC37G, through the Long Term Space Astrophysics
program. This research was performed in the Rochester Imaging
Detector Laboratory with support from a NYSTAR Faculty Development
Program grant. The work is also supported by a special grant from
National Natural Science Foundation of China, NSFC-10843008. Part of
the data presented here was obtained with the ESO New Technology
Telescope (PI: Benjamin Davies, PID: 080.D-0470). This publication
makes use of data products from the Two Micron All Sky Survey, which
is a joint project of the University of Massachusetts and the
Infrared Processing and Analysis Center, California Institute of
Technology, funded by the National Aeronautics and Space
Administration and the NSF. This research has made use of Spitzer's
IRAC imaging data (PI: Donald Figer, PID:30734), superCOSMOS's
H$_\alpha$ line sky survey and Spitzer's massive star cluster
servery and galactic structure and star-forming region imaging data
(PI: Steven R Majewski, PID:40791), PMN radio 3842 MHz sky survery
data, the SIMBAD database, and the GSFC IDL library. We also thank
Dr. Lise Deharveng for advices on radio observations.

\begin{deluxetable}{cccc}
\tabletypesize{\scriptsize}
\tablewidth{0pt}
\tablecaption{Magnitude Calibration Coefficients \label{tab_cali}}
\tablehead{
\colhead{Band}& 
\colhead{a(A)}&
\colhead{b(B)}&
\colhead{c(C)}
}
\startdata
 h &  2.396 &  1.008 &  0.007\\
ks &  2.876 &  0.001 &  1.019\\
 H & -2.359 &  0.992 & -0.007\\
Ks & -2.822 & -0.001 &  0.982\\
\enddata
\tablecomments{Magnitude transformation coefficients derived through the regression between 2MASS magnitudes and instrumental magnitudes.
}
\end{deluxetable}

\begin{deluxetable}{ccc}
\tabletypesize{\scriptsize}
\tablewidth{0pt}
\tablecaption{LIST OF STARS \label{tab_coor}}
\tablehead{
\colhead{Star\#}& 
\colhead{RA(J2000)}&
\colhead{DEC(J2000)}
}
\startdata
 1 & 154 ~ 48 ~00.5 ~ & -  58 ~ 02 ~23.4 ~\\
 2 & 154 ~ 47 ~31.7 ~ & -  58 ~ 02 ~26.6 ~\\
 3 & 154 ~ 47 ~15.6 ~ & -  58 ~ 02 ~34.6 ~\\
 4 & 154 ~ 47 ~47.8 ~ & -  58 ~ 02 ~13.0 ~\\
 5 & 154 ~ 47 ~13.8 ~ & -  58 ~ 02 ~13.9 ~\\
 6 & 154 ~ 47 ~35.7 ~ & -  58 ~ 01 ~59.5 ~\\
 7 & 154 ~ 47 ~26.5 ~ & -  58 ~ 01 ~55.5 ~\\
 8 & 154 ~ 46 ~39.9 ~ & -  58 ~ 01 ~16.0 ~\\
\enddata
\tablecomments{Coordinates of the stars indicated in Figure~\ref{fig_spec}.
}
\end{deluxetable}

\end{document}